\documentclass[12pt,preprint]{aastex}
\usepackage{amsmath}

\begin{document}

\title{Could SXP 1062 be an Accreting Magnetar?}
\author{
Lei Fu\altaffilmark{1,2} and Xiang-Dong Li\altaffilmark{1,2}}

\affil{$^{1}$Department of Astronomy, Nanjing University,
Nanjing 210093, China}

\affil{$^{2}$Key laboratory of Modern Astronomy and Astrophysics,
Ministry of Education, Nanjing 210093, China}

\begin{abstract}
In this work we explore the possible evolutionary track of the
neutron star in the newly discovered Be/X-ray binary SXP 1062, which
is believed to be the first X-ray pulsar associated with a supernova
remnant. Although no cyclotron feature has been detected to indicate
the strength of the neutron star's magnetic field, we show that it
may be $\ga 10^{14}\,$G. If so SXP 1062 may belong to the accreting
magnetars in binary systems. We attempt to reconcile the short age
and long spin period of the pulsar taking account of different
initial parameters and spin-down mechanisms of the neutron star. Our
calculated results show that, to spin down to a period $\sim
1000\,$s within $10-40\,$kyr requires efficient propeller
mechanisms. In particular, the model for angular momentum loss under
energy conservation seems to be ruled out.
\end{abstract}

\keywords{stars: neutron -- X-rays: binaries: SXP 1062}

\section{Introduction}

As a subgroup of high-mass X-ray binaries (HMXBs), Be/X-ray binaries (BeXBs)
consist of a neutron star (NS)  and a Be companion star which
show emission lines and infrared (IR) excess in its spectrum. The origin of the
emission lines and the IR excess is attributed to the circumstellar discs,
which is fed from the material expelled from the rapidly rotating Be stars.
The X-ray emissions is believed to originate from accretion of
matter in the circumstellar discs by the NSs \citep[see][for a review]{rei11}.

BeXBs are subdivided into persistent and transient sources according to their different
X-ray properties. Transient systems are characterized by  outbursting
activities, in which the X-ray flux increases by $\sim 1-4$ orders of magnitude
compared with in
quiescent state, and the outbursts typically last about $0.2-0.3$ orbital period.
These systems often have moderately eccentric ($e\ga 0.3$) and relatively narrow orbits
($P_{\rm orb} \la 100 \,{\rm days}$).
Persistent sources are relatively quiet systems with low X-ray luminosities
($L_{\rm X} \sim 10^{34}-10^{35}\,{\rm erg\,s^{-1}}$), the variability of which is less than
an order of magnitude. They usually have slowly rotating NSs ($P \ga 200\,{\rm s}$),
low eccentric ($e \la 0.2$) and relatively wide orbits ($P_{\rm orb} \ga 200
\, {\rm days}) $ \citep{rei11}.

Recently \citet{hen12} reported a new BeXB SXP 1062
in the Wing of the Small Magellanic Cloud (SMC).
This source was first discovered as a transient BeXB during the {\em XMM-Newton}
and {\em Chandra} observations in 2010, and was not active during the {\em ROSAT}
and {\em ASCA} observations of the SMC \citep{hab00,yok03}.
However, SXP 1062 seems to share some
characteristics with persistent BeXBs:
a relatively low intrinsic X-ray luminosity
$L_{\rm X} \simeq 6.3(^{+0.7}_{-0.8})\times 10^{35} \, {\rm erg\, s^{-1}}$
(corresponding an accretion rate of
$\dot{M}=L_{\rm X}/\eta c^2\sim 6\times 10^{15}\,{\rm g\,s^{-1}}$
with energy conversion efficiency $\eta=0.1$, $c$ is the speed of light),
a slowly rotating NS with period of $P\simeq 1062 \, {\rm s}$,
a relatively flat light curve with sporadic fluctuation less than
an order of magnitude, and a probably wide orbital period
$P_{\rm orb}\sim 300\,{\rm days}$ derived from the \citet{cor84} diagram.
What makes this discovery most noticeable is that SXP 1062 is located
in the center of a shell-like nebula, which is considered to be
a supernova remnant (SNR), aging only $\sim 10-40$ kyr \citep{hab12,hen12}.
Thus SXP 1062 provides the first example of an X-ray pulsar associated with a SNR,
and it challenges the traditional spin-down model of NSs because of its
extraordinarily long spin period combined with a relatively young age.

\citet{hab12} measured the spin period change in SXP 1062 over a 18
day duration of observation. Their timing analysis shows that the NS
in SXP 1062 has a very large average spin-down rate with the spin
frequency derivative  $\dot{\nu}\sim 2.6 \times 10^{-12}\,{\rm
Hz\,s^{-1}}$(or period derivative $\dot{P}\sim 100\rm\,s\,yr^{-1}$).
If the NS has a normal magnetic field ($B\sim 10^{12}-10^{13}$ G),
it's hard to spin-down to a period $\sim 1000$\,s within a few
$10^{4}$ years. Assuming in the extreme case that the NS has spun
down with a constant spin frequency derivative $-2.6 \times
10^{-12}\,{\rm Hz\,s^{-1}}$ over it's whole lifetime, \citet{hab12}
derived the lower limit of the initial spin period of the NS is 0.5
s. Since the duration of the observation lasts only 18 days
(probable less than one tenth of the orbit period), the
extraordinarily large spin-down rate is very unlikely to sustain in
the whole lifetime of SXP 1062, thus the present value of spin-down
rate may be just a short-term one.

\citet{pop12} (hereafter PT12) suggest another possibility
to reconcile the long spin period and short age of SXP 1062.
Assuming that the NS is spinning at the equilibrium
period, PT12 estimated the {\em current} magnetic field to be
$B\la 10^{13}$ G according to the model of \citet{sha12}.
Their calculations  show that if the NS in SXP 1062 was born as a magnetar
($B>10^{14}$ G) it can be
spun down to $\sim 1000$ s within a few $10^4$ yr.

In this work we consider the proposed mechanisms that can account
for the observed rapid spin-down in SXP 1062, if it is alternatively
a magnetar with current field strength $\ga 10^{14}$ G. In Section 3
we point out that this possibility remains according to current
observations. Based on this, we investigate its spin-down evolution,
taking account of various kinds of braking torques during the
propeller stage, which is introduced in Section 2, to examine how
the observations of SXP 1062 can constrain the possible spin-down
mechanisms in NSs. We present our calculated results in Section 4,
and discuss their possible implications in Section 5. We conclude
that the spin-down evolution is sensitive to the specific propeller
mechanism rather the initial spin period of the NS.

\section{Spin-down models of NSs in HMXBs}

\subsection{The ejector phase}
Normally a newborn NS in a binary first appears as a radio pulsar
(or ejector)
after the supernova explosion \citep{lip92}.
In this phase the spin-down of the NS is due to the loss of its rotational
energy dominated
by magneto-dipole radiation and the outgoing flux of relativistic particles.
If the NS's companion is a high-mass star, the stellar wind matter from the
companion within the gravitational radius of
$R_{\rm G}=2GM/{V^2}$ will be captured by the NS at a rate
$\dot{M}=\pi R^2_{\rm G} \rho V$. Here $G$ is the gravitational
constant, $M$ the mass of the NS, $\rho$ the density of the wind
at $R_{\rm G}$, and $V$ the velocity of the NS relative to
the stellar wind, i.e. $V=(V_{\rm orb}^2+V_{\rm w}^2)^{1/2}$,
where $V_{\rm orb}$ and $V_{\rm w}$ are the orbital velocity of the NS and the
wind velocity, respectively.
The pressure of the outgoing radiation and particles is
larger than that of the incoming matter
at $R_{\rm G}$.
The energy loss rate can be expressed as
$\dot{E}=-{\mu}^2 {\Omega}^4(1+\sin^2 \alpha)/{c^3}$
\citep{spi06}, where $\mu\equiv BR^3$
is the magnetic moment of the NS ($B$ and $R$ are the
surface magnetic field and radius of the NS, respectively), $\Omega$  the angular velocity,
and $\alpha$ the inclination angle between the  magnetic
and rotational axes. Thus the spin-down rate in the
ejector phase is
\begin{equation}
\dot{P}={4{\pi}^2 B^2 {R^6}(1+\sin^2 \alpha) \over IPc^3}\, ,
\end{equation}
where $I$ is moment of inertia of the NS.
As the NS spins down and the outgoing pressure goes down, the transition
to the supersonic propeller phase will occur when the two pressures are in balance.
The spin period of the NS at the transition point is
\begin{equation}
P_{\rm ej}={2\pi \over c} \left[ {B^2 R^6(1+{\sin}^2\alpha) \over 4 \dot{M} V} \right]^{1/4}.
\end{equation}

\subsection{The propeller phase}

Once the wind matter crosses the gravitational radius $R_{\rm G}$ the
propeller phase starts.
If the plasma enters the light cylinderical radius the pulsar mechanism will switch off
and the incoming matter will form a quasi-static atmosphere surrounding the NS.
At this moment accretion does not occur since the magnetosphere radius
$R_{\rm m}=(\mu^2/\dot{M}\sqrt{2GM})^{2/7}$
is larger than the corotation radius $R_{\rm co}\equiv (GM/{\Omega}^2)^{1/3}$
of the NS. The infalling material is stopped at the magnetosphere by
the centrifugal barrier, which prevents material from accreting onto the NS.
The ejected material will carry away the angular momentum of the
NS and decelerate its spin. This so-called propeller effect
was first introduced by \citet{ill75}.

\citet{dav79} pointed out that, according to the Mach number
${\mathcal{M}} = \Omega R_{\rm m}/c_{\rm s}$
(here $c_{\rm s}\sim (GM/ R_{\rm m})^{1/2}$ is the sound velocity at $R_{\rm m}$)
at the magnetosphere, the
propeller phase can be subdivided into two cases:
supersonic propeller and subsonic propeller.
Accordingly the above mentioned propeller mechanism
is related to the supersonic propeller since the Mach number ${\mathcal{M}} >1$.
This phase ends when $R_{\rm co} = R_{\rm m}$ (i.e., ${\mathcal{M}}=1$)
and the corresponding spin period
\begin{equation}
P_{\rm eq}=2^{11/14}\pi \mu^{6/7}\dot{M}^{-3/7} (GM)^{-5/7}\,
\end{equation}
is called the equilibrium period. Further works \citep{aro76,els77}
showed that, unless the material outside the magnetosphere is able to cool,
accretion is unlikely to happen. Thus, even with $R_{\rm co} > R_{\rm m}$
the propeller stage will succeed once the energy deposition rate is larger
than the energy loss rate of the surrounding shell, and keep removing
angular momentum from the NS.
Because the Mach number ${\mathcal{M}} <1$, this stage is called
the subsonic propeller.
This process will cease if the loss rate of the rotational energy
can no longer support the surrounding atmosphere against cooling,
then the atmosphere will collapse and the NS enters the accretor stage.
The spin period of the NS at this point is so-called the break period,
given by \citep{dav81,ikh01}
\begin{equation}
P_{\rm br}\simeq 86.88\,\mu_{30}^{16/21}\dot{M}_{16}^{-5/7}(M/M_{\odot})^{-4/21} \,\rm s\,,
\end{equation}
where $\mu_{30}=\mu/10^{30}\,{\rm G\,cm^3}$, and
$\dot{M}_{16}=\dot{M}/10^{16}\,{\rm g\,s^{-1}}$.

It should be noted that the supersonic propeller can occur in both
wind-fed and disc-fed cases. However, there is no consensus on the
angular momentum loss rate of a NS during the propeller phase
\citep{dav79}. Here we adopt a general formulation of the spin-down
torque as follows \citep{mor03},
\begin{equation}
I\dot{\Omega} =-\dot{M} R_{\rm m}^2 \Omega_{\rm K}(R_{\rm m})
\left[ {\Omega \over \Omega_{\rm K}(R_{\rm m})}\right]^{\gamma}\,,
\end{equation}
where $\gamma$ is a parameter ranging from $-1$ to 2, and its value
reflects various propeller mechanisms and spin-down efficiencies.
For the supersonic propeller, $\gamma$ =$-1$, 0 and 1.
When $\gamma=-1$, the matter is assumed to be ejected with the  escape velocity
at $R_{\rm m}$, i.e., $v_{\rm esc}(R_{\rm m})=\sqrt{2GM/R_{\rm m}}$ \citep{ill75},
and the spin-down torque is calculated based on energy budget.
The energy loss rate is $I\Omega\dot{\Omega}=-(1/2)\dot{M}v^2_{\rm esc}(R_{\rm m})
=-\dot{M}[R_{\rm m}\Omega_{\rm K}(R_{\rm m})]^2$, where $\Omega_{\rm K}(R_{\rm m})$
is the Keplerian angular velocity at $R_{\rm m}$.
When $\gamma=0$ and 1, the matter is assumed to be ejected at
the escape velocity $v_{\rm esc}(R_{\rm m})$ \citep{dav73} and the rotating velocity
$R_{\rm m}\Omega$ \citep{sha75}
of the magnetosphere at $R_{\rm m}$, respectively,
and the torque is derived under the angular momentum budget.
The corresponding angular momentum loss rate is
$I\dot{\Omega}=-\dot{M}R_{\rm m}(2GM/R_{\rm m})^{1/2}
 =-2^{1/2}\dot{M}R_{\rm m}^2 \Omega_{\rm K}(R_{\rm m})$ and
$I\dot{\Omega}=-\dot{M}R^2_{\rm m}\Omega$, respectively.
The value of $\gamma$ for the subsonic propeller phase
is 2, in which the rotational energy of the NS is assumed to be dissipated
at a rate of $I\Omega\dot{\Omega}=
-\dot{M}R_{\rm m}^2\Omega^2_{\rm K}(R_{\rm m})[\Omega/\Omega_{\rm K}(R_{\rm m})]^3$,
and the resulting torque is
$I\dot{\Omega}=-\dot{M} R_{\rm m}^2 \Omega_{\rm K}(R_{\rm m})
[{\Omega/\Omega_{\rm K}(R_{\rm m})}]^2$. Thus
the spin-down rate in the propeller stage can be summarized as
\begin{equation}
\dot{P}={{(2\pi)^{\gamma-1}(GM)^{1-\gamma \over 2}
\dot{M}R_{\rm m}^{1+3\gamma \over 2}}\over {I P^{\gamma -2}}}.
\end{equation}
As an illustration, we consider a $1.4\,M_{\sun}$ NS with an initial spin period of
0.01 s, a surface magnetic field of $10^{12}\,$G, and an accreting rate of
$10^{16}$ gs$^{-1}$.  The corresponding characteristic spin-periods
are $P_{\rm ej}\simeq 0.24\,$s, $P_{\rm eq}\simeq 6.7\,$s and $P_{\rm br}\simeq 81.5\,$s
respectively. The timescales in the supersonic propeller phase varies from 30\,kyr to 20\,Myr
as $\gamma$ decreases from 1 to $-1$, and in the subsonic propeller phase which has a spin-down
rate irrelevant with $P$ the spin-down timescale is $\sim 90\,$kyr. In
the above calculation we use
$300\,{\rm km\,s^{-1}}$ as the relative velocity \citep[see][]{rag98}.

\subsection{The accretor phase}

Steady wind accretion onto the NS starts at $P>P_{\rm br}$. In this phase
the spin period could be further changed since the wind matter possesses some
angular momentum. However both observations \citep{bil97} and numerical calculations
\citep[e.g.,][]{fry88,mat92,anz95,ruf99}
have shown that the efficiency of angular momentum transfer
in wind accretion is quite low, with alternating short-term spin-up and
spin-down. Thus one may expect that the present spin periods of
wind-fed X-ray pulsars are not significantly different from
the $P_{\rm br}$ achieved earlier.

Recently \citet{sha12} proposed a model of subsonic quasi-spherical
accretion onto a slowly rotating NS in HMXBs
with low X-ray luminosities ($L_{\rm X}<10^{36}\,{\rm erg\,s^{-1}}$).
In this model the accreting matter settles down  subsonically onto the rotating
magnetosphere, forming an extended quasi-static shell around it.
The angular momentum can be removed from or injected into the NS
depending on the sign of the specific angular momentum of the falling matter.
In the case of moderate coupling between the plasma and the magnetosphere,
from the torques acted on the NS due to both the magnetosphere-plasma
interaction and accretion,
the changing rate of the spin period is derived to be
(see also PT12)
\begin{equation}
\dot{P}=-{P^2 \over 2\pi I}\left[ A\dot{M}^{(3+2n)/11}_{16} - C\dot{M}^{3/11}_{16} \right]\,,
\end{equation}
where
$A\sim 2.2\times 10^{32} K_1(B_{12}R_{6})^{1/11}V_{300}^{-4}P_{{\rm orb},300}^{-1}$,
and $C\sim 5.4\times 10^{31} K_1(B_{12}R_{6})^{13/11} P_{1000}^{-1}$.
Here $B_{12}=B/10^{12}\,{\rm G}$, $R_{6}=R/10^6\,{\rm cm}$,
$P_{1000}=P/1000\,{\rm s}$, $P_{\rm orb,300}=P_{\rm orb}/300\,{\rm hr}$,
$V_{300}=V/300\,{\rm km\,s^{-1}}$.
The constants $K_1$ and $n$ are set to be 40 and 2 respectively.
It is seen that there is an critical accretion rate at which $\dot{P}=0$.

\section{Estimate of the magnetic field}

The NS magnetic field is a critical parameter in the spin-down
models. Before investigating the spin history of SXP 1062, we need
to know the information about its magnetic field strength. The
cyclotron features in the X-ray spectra provide the most direct and
accurate way to measure the magnetic field strengths of accreting
NSs. Unfortunately, they have not been detected in SXP 1062.
Nevertheless, there are several other ways to estimate the NS
magnetic field from its spin period and period derivative, though
model dependent.

One of the hints comes from the young age of the SNR associated with the NS.
This requires that the lifetime of the ejector phase (usually much longer than that of
the propeller phase) must end within a few $10^4$ years.
Assuming that the magnetic field has changed little during this phase and that the
initial spin period is much smaller than $P_{\rm ej}$, one can estimate the
timescale of the ejector phase to be (PT12)
\begin{equation}
\tau_{\rm ej}= {c^3IP_{\rm ej}^2\over16\pi^2B^2R^6}\sim 1.5\,
\dot{M}_{16}^{-1/2}V_{300}^{-1/2}B_{12}^{-1}\,{\rm Myr}.
\end{equation}
This value is about two orders of magnitude larger than
the estimated age of SXP 1062, unless $B_{12}>100$. This means that SXP 1062
must have possessed very strong magnetic field.

PT12 further assumed that the NS in SXP 1062 is spinning at the
equilibrium period as described in the model of \citet{sha12}, and
derived the current magnetic field to be $B_{12}\la 10^{13}$ G using
Eq.~(7). Accordingly they suggest that the NS magnetic field must
have been stronger in the past and then decayed to its present,
normal value. It is noted that the model of \citet{sha12} has quite
a few parameters whose magnitudes are uncertain. For example, the
value of $K_1$, which relates the poloidal ($B_p$) and toroidal
($B_{\phi}$) magnetic field components, is found to be $\sim 40$ in
\citet{sha12}. This will result in $B_{\phi}\gg B_p$ during the
accretor phase, and it is not known whether the magnetic field
configuration can remain stable in this case
\citep[cf.][]{aly85,wan95}.

The extraordinary large spin-down rate of SXP 1062 can be used to put useful
constraint on the magnetic field of the NS. As shown by many authors \citep{lyn74,lip82,bis91},
the maximum spin-down torque exerted on a NS in either disc or spherical accretion is
\begin{equation}
I\dot{\Omega}=-\kappa{\mu^2\over R_{\rm co}^3},
\end{equation}
where $\kappa< 1$.
To account for the spin-down rate measured in SXP 1062, the
NS magnetic field has to be
\begin{equation}
B\simeq 3\times 10^{14}\kappa^{-1/2}M_{1.4}^{1/2}I_{45}^{1/2}
R_6^{-3}(\dot{P}/100\,{\rm syr}^{-1})^{1/2}\,{\rm G},
\end{equation}
where $M_{1.4}=M/1.4M_{\sun}$, and $I_{45}=I/10^{45}$ gcm$^2$. The same result
can be obtained if the spin-down torque in the subsonic propeller phase
\citep{dav81} is used.

Another efficient spin-down mechanism was proposed by \citet{ill90}.
They argued that there could be outflows from the NS magnetosphere
caused by heating of hard X-ray
emission of the NS if the X-ray luminosity falls in the range of
$\sim 2 \times10^{34}$ ergs$^{-1} - 3\times 10^{36}$ ergs$^{-1}$.
Compton scattering heats the accreted matter anisotropically, and some of the heated
matter with a low density can flow up and form outflows to take
the angular momentum away. The corresponding spin-down torque is
\begin{equation}
I\dot{\Omega}=-\kappa{\chi\over 2\pi}\dot{M}_{\rm out}\Omega R_{\rm m}.
\end{equation}
Here $\dot{M}_{\rm out}$ is the mass outflow rate (no larger than the mass transfer rate)
and $\chi$ is the solid angle of the outflow. This
gives the magnetic field to be
\begin{equation}
B\simeq 3.6\times 10^{14}({\kappa\chi\over 2\pi})^{-7/8}I_{45}^{7/8}M_{1.4}^{1/4}I_{45}^{1/2}
R_6^{-3}({\dot{M}_{\rm out}\over 10^{16}\,{\rm gs}^{-1}})^{-3/8}
({\dot{P}\over 100\,{\rm syr}^{-1}})^{7/8}({P\over 1062\,{\rm s}})^{-7/8}\,{\rm G}.
\end{equation}

The above estimates show that SXP 1062 could be an accreting magnetar. Similar
conclusions have also been drawn for other X-ray pulsars in HMXBs.
\citet{dor10} reported the spin history of the 685 s X-ray pulsar GX 301$-$2,
and found it spinning down at a rate $\dot{\nu}\sim 10^{-13}$ Hzs$^{-1}$.
Reig et al. (2012) showed that the measurements of the spin period (5560 s) of 4U2206$+$54
imply a spin-down rate of $\dot{\nu}\sim -1.5(\pm 0.2)\times 10^{-14}$ Hzs$^{-1}$.
Using the above spin-down mechanisms to explain the spin-down rates also leads to
very strong magnetic fields ($>10^{14}$ G) in these NSs  \citep[see also][]{lip82}.

\citet{if12} suggested an alternative interpretation
for the rapid spin-down in GX 301$-$2.
They showed that if the accreting material is magnetized, the magnetic
pressure in the accretion flow increases more rapidly than its ram pressure, and under
certain conditions the magnetospheric radius
\begin{equation}
R_{\rm mca}\simeq 1.5\times 10^{8}\alpha_{0.1}^{2/3}B_{12}^{6/13}R_6^{18/13}
T_6^{-2/13}M_{1.4}^{1/13}\dot{M}_{16}^{-4/13}\,{\rm cm},
\end{equation}
is considerably smaller than the traditional magnetospheric radius. Here $\alpha=0.1
\alpha_{0.1}$ is the efficiency parameter of Bohm diffusion, and $T=10^6T_6$ K
is the plasma temperature at the magnetospheric boundary.
The spin-down torque applied to the NS is found to be
\begin{equation}
I\dot{\Omega}=-\frac{\kappa_m\mu^2}{(R_{\rm co}R_{\rm mca})^{3/2}},
\end{equation}
where $\kappa_m$ is a dimensionless efficiency parameter for the magnetic viscosity
coefficient, and $0<\kappa_m<1$. The above equation can explain
the spin-down of  GX 301$-$2 with a normal field of a few $10^{12}$ G
{\bf if $\kappa_m\sim 0.1$.}
In the case of SXP 1062, it
results in the estimate of the magnetic field to be
\begin{equation}
B\simeq 2\times 10^{14}\kappa_{0.1}^{-13/17}I_{45}^{13/17}M_{1.4}^{8/17}I_{45}^{13/17}
R_6^{-3}\dot{M}_{16}^{-6/17}T_6^{-3/17}
({\dot{P}\over 100\,{\rm syr}^{-1}})^{13/17}({P\over 1062\,{\rm s}})^{-13/17}\,{\rm G},
\end{equation}
where $\kappa_{0.1}=\kappa_{m}/0.1$. In the same way, \citet{ikh12}
estimated the magnetic field of SXP 1062 to be $\sim 4\times
10^{13}$ G by assuming $\kappa_m=1$. Even this limiting value is
comparable to the quantum critical field $B_{\rm Q}=4.4\times
10^{13}$ G.

According to the above arguments, in the following we assume that the current
magnetic field of SXP 1062 is $\ga 10^{14}$ G. As to the evolution of the
magnetic field, we consider two kinds of models. First we assume that
the magnetic field was initially stronger and adopt a
phenomenological model for the magnetic field decay \citep{dall12}
\begin{equation}
\frac{{\rm d} B}{{\rm d} t}=-AB^{1+\alpha}=-\frac{B}{\tau_{\rm d}(B)},
\end{equation}
where the field decay timescale $\tau_{\rm d}(B)=(AB^\alpha)^{-1}$, and
$A$ and $\alpha$ are constants. The solution of the above equation in the case of $\alpha
\neq 0$ is
\begin{equation}
B=B_{\rm i}(1+\alpha t/\tau_{\rm d,i})^{-1/\alpha},
\end{equation}
where $B_{\rm i}$ is the initial field strength and $\tau_{\rm d,i}
=(AB_{\rm i}^\alpha)^{-1}$. \citet{dall12} show that, to be compatible with
the observations of magnetar candidates, the magnetic field should decay
on a timescale of $\sim 10^3$ yr for $B\sim 10^{15}$ G, with a decay index
most likely within the range $1.5\la \alpha \la 1.8$. Here we adopt
the initial magnetic field as $7\times 10^{14}$, $3\times 10^{14}$ and
$10^{14}\, {\rm G}$, with $\alpha=1.6$ and $\tau_{\rm d,i}=10^3/B_{\rm i,15}^{\alpha}$
yr where $B_{\rm i, 15}=B_{\rm i}/10^{15}$ G.

On the other hand, the observed braking indices for several young radio pulsars
have been measured and are all less than 3
\citep{lyne93,lyne96,kas94,liv05,liv06,liv11,wel11}, suggesting that the NS
magnetic fields may be increasing.
In particular, the braking index of the high-field ($5\times 10^{13}$ G) radio pulsar
PSR J1734$-$3333 was found to be $0.9\pm 0.2$ \citep{esp11},
implying that this pulsar may soon have the rotational properties of a magnetar.
In the second approach,
we adopt a field growth model in the following form
\begin{equation}
B=B_{\rm i}(1+t/\tau)^{\alpha},
\end{equation}
with $B_{\rm i}=3\times 10^{12}$ G, $\tau=10^3$ yr, and $\alpha=1.45$,
so that $B=8.5\times 10^{13}$ G and $6.3\times 10^{14}$ G at $t=10^4$ and
$4\times 10^4$ yr, respectively.

In Figure 1 we show the model evolution of the magnetic fields.

\section{Spin evolution}

A newborn NS is usually rotating rapidly. However, \citet{hab12} suggested that SXP 1062
could have been born with a period much larger
than $0.01$\,s. Some central compact objects (CCOs) in supernova
remnants which have spin periods ranging from $\sim 0.1$ to $\sim 0.5$\,s
\citep{zav00,g05,gh09} seem to  support this point of view, since there is evidence that
the spin periods of these sources are very close to the initial ones.
Thus in our model we take 0.01\,s, 0.5\,s and 6.5\,s as the initial period of the NS
in order to examine whether it can significantly influence the spin-down evolution.
We use the ultra-long initial spin period of 6.5\,s because this value is larger than
$P_{\rm ej}$ with $B=7 \times 10^{14}\,$G in the ejector phase,
so that the NS will directly enter the supersonic
propeller phase after the SN event.

It was shown by \citet{aro76} and \citet{els77} that, for
stable accretion to occur, the plasma at the base of the NS magnetosphere
should become sufficiently cool, so that the magnetospheric
boundary is unstable with respect to interchange instabilities. This can be realized
only if the spin period of the star exceeds the break period $P_{\rm br}$,
and the X-ray luminosity is larger than
\begin{equation}
L_{\rm cr}=3\times 10^{36}B_{12}^{1/4} M_{1.4}^{1/2}R_6^{-1/8} \,{\rm ergs}^{-1}.
\end{equation}
If $B\ga 10^{14}$ G, SXP 1062 should be in the subsonic propeller
phase. However, it is not clear whether the picture of the subsonic
propeller can be applied to BeXBs, due to the following reasons. (1)
The mass accretion in BeXBs is now believed to be triggered by
Roche-lobe overflow of the Be discs which is truncated by the NS
through a tidal torque \citep{oka01,oka02,rei11}, thus is deviated
from the traditional Bondi accretion in supergiant HMXBs. This means
that the NS in SXP 1062 is probably surrounded by a (quasi-)disc
rather a quasi-static, spherical atmosphere. (2) Even in the
spherical wind-fed case, the spin period - orbital period
correlation in BeXBs seems to be well accounted for by assuming that
the NSs are spinning at the equilibrium periods described by Eq.~(3)
\citep{cor84,wat89}\footnote{ Additionally, population synthesis
calculations by \citet{dai06} showed that the distribution of the
spin and orbital periods of X-ray pulsars in supergiant HMXBs can be
roughly explained without requiring that an X-ray pulsar emerges
after the subsonic propeller phase \citep[see also][]{ste86}.}. Thus
in our calculations we don't consider the subsonic propeller phase,
and assume that the evolutionary sequence of the NS is
ejector-supersonic propeller-accretor. We use different $\gamma$ to
calculate the spin-down torque in the supersonic propeller phase,
and take the equilibrium period (Eq.~[3]) as the final period (i.e.,
the period does not change during the accretor phase unless the
magnetic field changes).

In Figures $2-4$ we show the calculated results corresponding to
different initial spin periods of the NS. Here we take
the NS mass to be $M=1.4 M_{\sun}$, the inclination angle $\alpha =90^{\circ}$,
$I=10^{45}\,\rm g\,s^{-1}$, and $R=10^{6}\rm\,cm$.
The relative wind velocity $V$ is set to be $300\,\rm km\,s^{-1}$, and
the accretion rate is fixed to be $10^{16}\,\rm g\,s^{-1}$.
The three thin lines (from top to bottom)  describe the spin evolution
with initial magnetic field
of $7\times 10^{14}$ G, $3\times 10^{14}$ G, and $10^{14}$ G
undergoing field decay, respectively;
the thick line is for the field growth model with initial field of
$3\times 10^{12}$ G. The solid, dashed, and dotted lines represent
the ejector, propeller, and accretor phases, respectively.

We notice that the time spent in the supersonic propeller phase is sensitive
to the value of $\gamma$ which reflect different spin-down mechanisms in the
supersonic propeller phase as we mentioned before.
In the case of $\gamma=-1$, where the spin-down torque is
most inefficient, the NS has not evolved out of the supersonic
propeller phase at the age of the SNR, even with a superstrong magnetic
field. Our results are not sensitive to the
initial spin period of the NS.
Thus even for the case of ultra-long initial spin period
there is no significant change in the final NS period.

In other cases the NS can successfully reach  $P_{\rm eq}$ when
$t=10-40$ kyr.
Since $P_{\rm eq}$ depends on the magnetic field, we can see that the magnetic field
determines the final spin period which the NS can achieve, and the value of $\gamma$ determines
the evolutionary timescale. In the field decay model, the spin period
remains invariant once it reaches $P_{\rm eq}$, since we assume
that during the accretor phase the long-term, net torque from the wind is small.
In the field growth model, the spin period keeps increasing with $P_{\rm eq}$
in the final stage, since the increase of $B$ always breaks the instaneous
equilibrium and causes a spin-down torque.
It is seen that $B\ga 10^{14}\,\rm G$ can fulfill the requirement
of SXP 1062 in both models. This result favors that in the supersonic
propeller phase matter is ejected at $R_{\rm m}$ with the escape velocity
under angular momentum conservation, consistent with the
numerical calculation by \citet{wan85}.

\section{Discussion and conclusion}

The newly found Be/X-ray binary SXP 1062 is believed to be the first
X-ray pulsar associated with a SNR, which shows a combination of
young age and long spin period that cannot be explained by a typical
NS. Previous studies \citep{hab12,pop12} explored its possible
origin invoking initially long spin period or ultra-strong magnetic
field. Here we discuss the possibility that SXP 1062 is an accreting
magnetar with $B\ga 10^{14}$ G, and examine in this case how the
properties of the NS (i.e. initial spin period, magnetic field and
its evolution) and the spin-down torques can be constrained.

Other candidates of accreting magnetars in binaries include 4U
2206$+$54 \citep{fin10,rei12} and GX 301$-$2 \citep{dor10}. However,
it is controversial whether they really possess ultra-strong
magnetic fields. \citet{wan10} reported the existence of two
cyclotron absorption lines at $\sim 30$ and 60 keV in  4U 2206$+$54,
and derived a magnetic field of $3.3\times 10^{12}$ G, although no
sign of this feature has been detected in other observations.
Observations of \citet{la05} showed the cyclotron resonance
scattering feature at $\sim 35-45$ keV in GX 301$-$2, suggesting the
field strength of $4\times 10^{12}$ G. \citet{if12} proposed a
magnetic wind accretion model for  GX 301$-$2 to account for the
difference in the field strengths derived from the cyclotron lines
and from the spin-down rates. In the case of SXP 1062, it is found
that the magnetic field may be at least strong as $\sim B_{\rm Q}$
in the model of \citet{ikh12}. For a dipole magnetic field of $\sim
10^{14}$ G, the electron cyclotron line would appear at $E > 1$ MeV,
but a proton cyclotron line would appear at $E \sim 0.5(B/10^{14}
{\rm G}) = 0.3$ keV. Although a line with this energy should be
observable with {\em XMM-Newton} detectors, it is in a region
affected by strong interstellar absorption \citep{rei12}. Currently,
no significant lines have been detected in the persistent emission
of magnetars \citep{mer08}.

If SXP 1062 is or has been a magnetar, the association between the
SNR and SXP 1062 may provide an opportunity to investigate the
formation and evolution of magnetars. \citet{vink06} showed that
there is no evidence that magnetars are formed from rapidly rotating
proto-neutron stars. The SNR associated with SXP 1062 is one of the
faintest SNRs known in the SMC \citep{fil05,fil08,owen11}. This
seems to in line with the finding of \citet{vink06} that their
formation may not be accompanied  with extraordinarily bright
supernovae. However, it is known that the brightness of SNRs depends
strongly on the density of the environment. Nevertheless, the age of
the SNR can be used to set useful constraints on the timescale of
magnetic field evolution, either due to field decay or growth.

Since the long spin period is most likely to be reached during the
propeller phase, the age of the SNR also plays a role in testing the
efficiency of the spin-down torques in different propeller
mechanisms. Our results seem to rule out the model with $\gamma=-1$,
and prefer larger values of $\gamma$ which correspond to more
efficient propeller spin-down. Recent 2- and 3-dimensional
magnetohydrodynamic (MHD) simulations by \citet{rom05} and
\citet{ust06} on disc-fed NSs suggest $\dot{\Omega}\propto
-\Omega^{2}$ for propeller-driven outflows. \citet{tor10}
investigate the spinning-down of magnetars rotating in the propeller
regime with axisymmetric MHD simulations, and find
$\dot{\Omega}\propto -\Omega^{1.5}$. It should be noted that the
mass transfer rate is assumed to be constant throughout our
calculations, but in reality it must have varied with the orbital
motion of the NS. For instance, an eccentric orbit may result in
alternation among the ejector, propeller and accretor phases. The
sporadic outburst behavior will further complicate the spin-down
evolution of the NS. This means that the calculated evolutionary
sequence in our model and the values of $\gamma$ should be taken as
an illustration and lower limits, respectively. However, both the
high spin-down rate and the young age of SXP 1062 provide strong
evidence that the binary indeed harbors or harbored a magnetar, and
a effective spin-down mechanism is required. We expect further
observations to confirm the long-term spin behavior of SXP 1062.

\begin{acknowledgements}

We thank an anonymous referee for helpful comments.
This work was supported by the Natural Science Foundation of China
under grant number 11133001 and the Ministry of Science, the
National Basic Research Program of China (973 Program 2009CB824800),
and the Qinglan project of Jiangsu Province.

\end{acknowledgements}

\clearpage

\begin{figure}
\epsscale{0.75}
\plotone{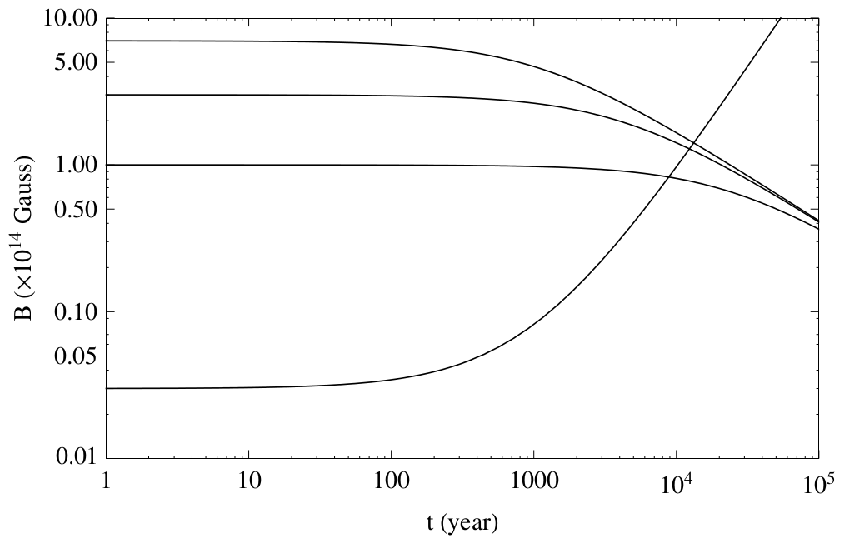}
\caption{Evolution of the NS magnetic fields in different models. \label{fig1}
}
\end{figure}

\begin{figure}
\epsscale{0.5}
\plotone{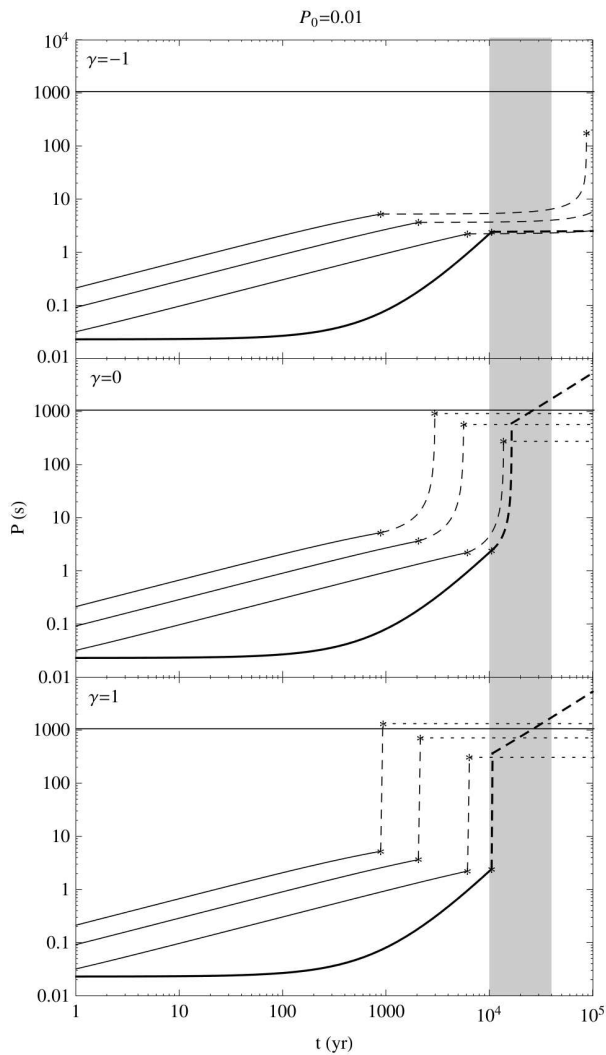}
\caption{The spin-down evolution for a NS with a 0.01\,s initial spin period.
In this and following figures, solid, dashed and dotted lines represent the ejector,
supersonic propeller and accretor phases, respectively.
Symbol * represents the transition of evolutionary stages.
The shaded area and the horizontal line mark the
age and the spin period of SXP 1062 respectively.
From the top to the bottom, the initial magnetic field is
$B_{\rm i}=7\times 10^{14}\,{\rm G}$, $3\times 10^{14}\,{\rm G}$,  $10^{14}\,{\rm G}$,
and $3\times 10^{12}\,{\rm G}$. \label{fig2}
}
\end{figure}

\clearpage

\begin{figure}
\epsscale{0.5}
\plotone{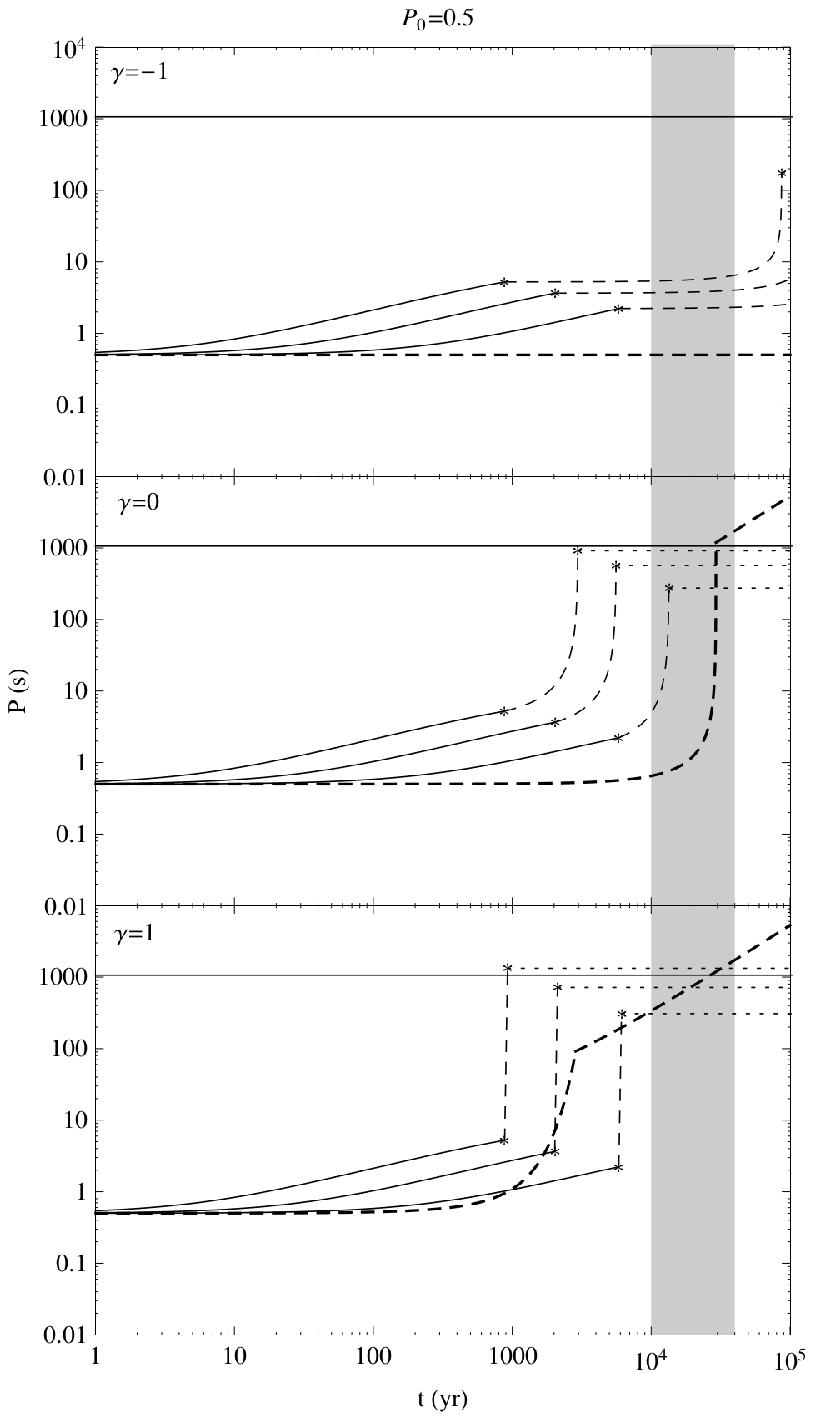}
\caption{The spin-down evolution for a NS with a 0.5\,s initial spin period.
Other parameters are similar to in Figure 2. \label{fig3}}
\end{figure}

\clearpage

\begin{figure}
\epsscale{0.5}
\plotone{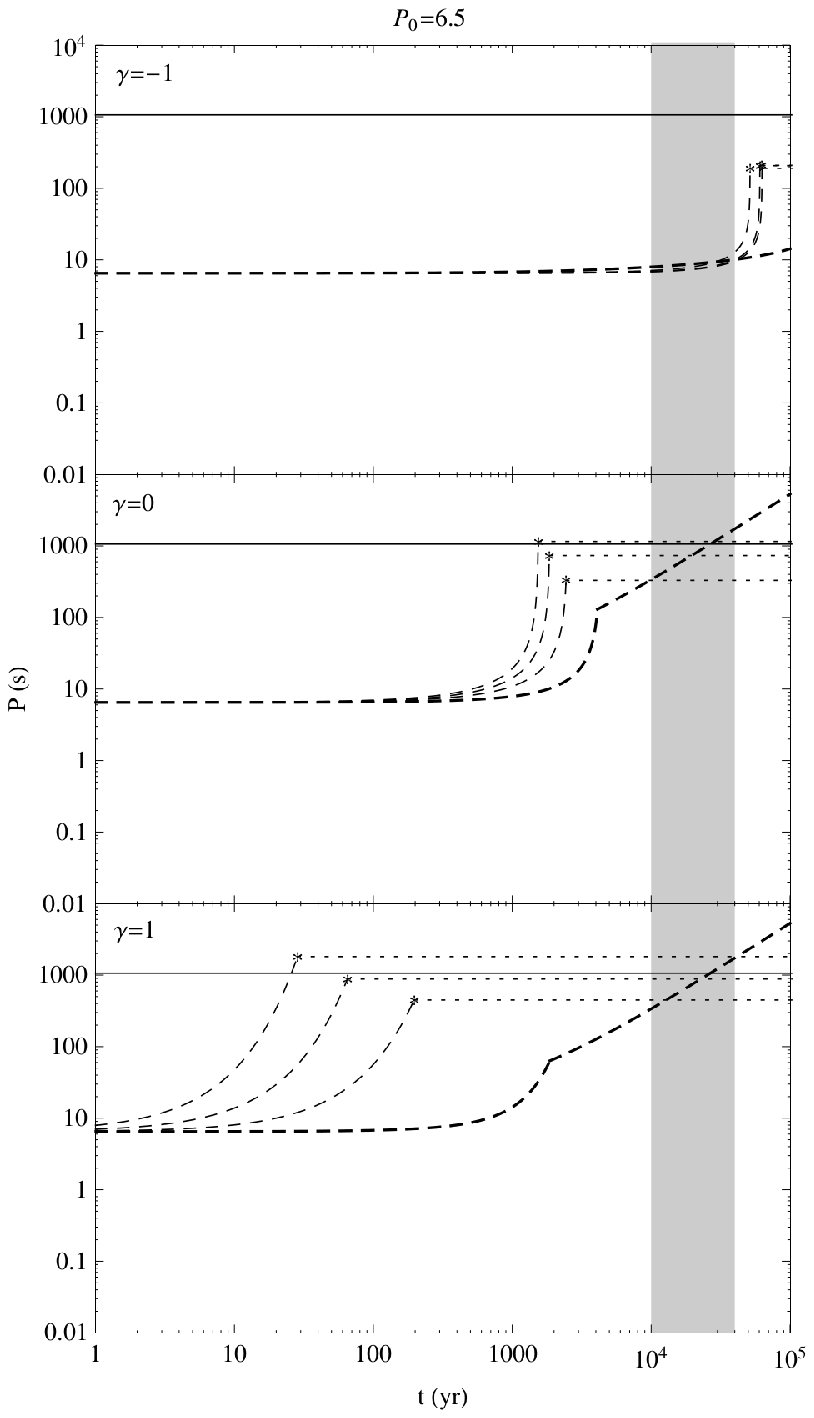}
\caption{The spin-down evolution for a NS with a 6.5\,s initial spin period.
Other parameters are similar to in Figure 2. \label{fig4}}
\end{figure}

\end{document}